\documentclass{appolb}
\usepackage{graphicx}
\usepackage[latin1]{inputenc}
\usepackage{amsmath}
\usepackage{amsfonts}
\usepackage{amssymb}
\usepackage{graphicx}
\usepackage{cite}
\usepackage{bm}
\usepackage{slashed}
\usepackage{hyperref}
\usepackage{float}
\usepackage{placeins}
\usepackage{flafter}
\usepackage{caption}
\usepackage{subcaption}
\graphicspath{{figures/}}

\begin{document}
\title{Published in Acta Phys. Pol. B 52, 1357 (2021) \\ \vspace{0.3in} \\
	Calculations of the alpha decay half-lives of some Polonium isotopes using the double folding model%
}
\author{W. A. Yahya \footnote{email: wasiu.yahya@gmail.com, wasiu.yahaya@kwasu.edu.ng }, K. J. Oyewumi$^*$
\address{$^\dagger$Department of Physics and Materials Science, Kwara State University, Malete, Kwara State, Nigeria
\\
$^*$Department of Physics, University of Ilorin, Nigeria} 
}
\maketitle
\begin{abstract}
The calculations of the alpha decay half-lives of some Polonium isotopes in the mass range $186-218$ have been carried out using the Wentzel-Kramers-Brillouin (WKB) semiclassical approximation. The alpha-nucleus effective potential used contains the Coulomb potential, centrifugal potential, and the nuclear potential. The nuclear potential is obtained via the double folding model, with the microscopic $\mathrm{NN}$ effective interactions derived from relativistic mean field theory Lagrangian (termed R3Y). Different parametrizations of the R3Y interactions have been employed in the computation of the nuclear potentials. The results obtained using the R3Y $\mathrm{NN}$ interactions are compared with the ones obtained using the famous  Michigan-3-Yukawa (M3Y) interactions. The use of density-dependent $\mathrm{NN}$ interaction is also considered. When compared to available experimental data, there are improvements in the results when density-dependent interaction potentials are used compared to when density-independent interactions are employed.
\end{abstract}
\PACS{27.90. +b; 23.60.+e ; 21.10.Tg ; 23.70. +j}
  
\section{Introduction}
Alpha decay is an important decay mode that can give information about the structure of nuclei \cite{shin2016,zanganah2020}. $\alpha-$decay of nuclei have been investigated using various theoretical approaches such as the generalised liquid drop model \cite{royer2001,xiaojun2014,royer2008}, the effective liquid drop model \cite{cui2018}, the modified generalized liquid drop model \cite{santhosh2018,santhosh2019,santhosh2020}, the fission-like model \cite{wang2010}, the preformed cluster model \cite{gupta1994,singh2010}, and cluster formation model \cite{deng2018,ahmediop2013,deng2016,ahmed2017}. These models use various interaction potentials ranging from the phenomenological potential such as the proximity potentials \cite{yahya2020}, the Woods-Saxon, squared Woods-Saxon, and Cosh potentials to microscopic interactions such as the double folding model. The Geiger-Nuttall law was the first decay law to describe $\alpha-$decay half-life, and Gamow in 1928 gave a theoretical explanation of the Geiger-Nuttall law. Gamow explained that the $\alpha-$decay was due to the quantum mechanical tunneling of a charged $\alpha$ particle through the nuclear Coulomb barrier \cite{zdeb2013}. Various empirical formulas have been introduced to compute the $\alpha$-decay half-lives of many isotopes since the introduction of the Geiger-Nuttall law. Some of these formulas are the Royer formula \cite{royer2000,royer2010,royer2011}, the Viola-Seaborg formula \cite{viola1966},  the  universal decay law developed by Qi et al. \cite{qi2009prc,qi2009prl},  the Akrawy formula \cite{akrawy2018}, the Ren formula, \cite{ren2004,akrawy2019}, the scaling law of Horoi \cite{horoi2004}, scaling law of Brown, the AKRE formula developed by Akrawy and Poenaru \cite{akrawy2017}, etc.\\

From a theoretical point of view, $\alpha-$decay half-lives can be studied using the semiclassical WKB framework. In this formalism, the effective interaction between the alpha-daughter system plays an important role in the calculations. The effective interaction consists of the nuclear potential, the Coulomb potential and the centrifugal potential. There have been various phenomenological \cite{naderi2013,adel2017} and microscopic nuclear potentials \cite{santhosh2012,ibrahim2017,soylu2012,soylu2018,ni2009} introduced to study the $\alpha-$decay of various nuclei. In the microscopic approach, the nuclear potential is determined using the double folding model, where the nuclear densities are folded with the effective M3Y nucleon-nucleon interaction. The use of density-dependent double folding model have also been introduced \cite{ni2009,xu2005,xu2006} to study the $\alpha-$decay half-lives of many nuclei. A microscopic $\mathrm{NN}$ interaction derived from relativistic mean field theory Lagrangian (termed R3Y) was introduced in Ref. \cite{singh2010} where the authors used the derived $\mathrm{NN}$ interaction to compute the optical potential in the double folding model and studied cluster decays of some nuclei. \\

In this study, the $\alpha-$decay half-lives of some Polonium isotopes have been calculated using both density-independent and density-dependent double folding model. The nuclear potential are calculated using the effective nucleon-nucleon interactions determined from relativistic mean field theory (termed R3Y). The results of the calculations using the M3Y-Paris and M3Y-Reid effective nucleon-nucleon interactions have also been included for comparison. The article is organised as follows: the theoretical models employed to compute the $\alpha$-decay half-lives of the Polonium isotopes are described in Section \ref{theory}. The results of the calculations are presented and discussed in Section \ref{results} while the conclusion is given in Section \ref{conclusion}.

\section{Theoretical Formalism}

\label{theory}

The effective alpha-nucleus potential $V(R)$ is given by
\begin{equation}
V_{eff}(R) = \lambda V_N(R)+V_C(R)+ V_{\ell},
\label{veff}
\end{equation}
where $\lambda$ is the quantization factor, $R$ is the relative distance between the alpha particle and daughter nucleus. The centrifugal term $V_{\ell} = \frac{\hbar^2 \ell (\ell+1)}{2 \mu R^2}$, $\ell$ is the orbital angular momentum, $\mu = m A_1 A_2/(A_1+A_2) $ is the reduced mass of the $\alpha$ particle and the daughter nucleus, and the nucleon mass $m = 931.494 \: \mathrm{MeV} $. By using Langer modification we have $\ell (\ell+1) \rightarrow \left( \ell+\frac{1}{2} \right)^2$. The values of $\ell$ are calculated by using the spin-parity selection rule \cite{maroufi2019polonica}:
\begin{equation}
\left| J_d - J_p \right| \leq \ell \leq J_d + J_p ,
\end{equation}
\begin{equation}
\pi_p = \left(-1\right)^\ell \pi_d.
\end{equation}
The Coulomb potential $V_C(R)$ is given in the form \cite{ni2009}
\begin{equation}
V_C (R) = Z_1 Z_2 e^2 \left\lbrace  \begin{array}{ll}
\frac{1}{R} & \textrm{ for } R > R_C \\ 
\frac{1}{2 R_C} \left[ 3 - \left( \frac{R}{R_C} \right)^2 \right] & \textrm{ for } R \leq R_C
\end{array}   \right. 
\end{equation}
where $Z_1$ and $Z_2$ are the charge number of the alpha particle and daughter nucleus, respectively, and $R_C = 1.2 \left( A_1^{1/3} + A_2^{1/3} \right)$. \\

The nuclear interaction potential $V_N(R)$ between the alpha and daughter nuclei in the double folding model is written as:
\begin{equation}
V_N(R) = \int \int \rho_1(\textbf{r}_1) F(\rho_1,\rho_2) \rho_2 (\textbf{r}_2) v(E_\alpha,s) d \textbf{r}_1 d \textbf{r}_2, 
\label{i1}
\end{equation}	
where $s = \left| R+\textbf{r}_2-\textbf{r}_1 \right|$ is the relative distance between interacting nucleon pair, $\rho_1(\textbf{r}_1)$ and $\rho_2(\textbf{r}_2)$ are the ground state matter density distributions of the alpha and daughter nuclei, respectively, and the kinetic energy of the $\alpha$ particle is denoted as $E_{\alpha}$ . The density distribution of the alpha particle is taken to be the usual Gaussian form:
\begin{equation}
\rho_1(r_1) = 0.4299 e^{-0.7024 r_1^2},
\end{equation}
and the density distribution of the daughter nucleus is taken to be the Fermi form \cite{xu2005,christel2018}:
\begin{equation}
\rho_2 (r_2) = \frac{\rho_0}{1+\exp \left( \frac{r_2-R}{a} \right)},
\end{equation}
where the diffuseness parameter $a = 0.54 \, \mathrm{fm}$, $R_{1(2)} = 1.07 A^{1/3}_{1(2)} (\mathrm{fm})$, $A_1$ is the mass number of the alpha particle and $A_2$ is the mass number of the daughter nucleus \cite{ghorbani2020,xu2005}. The value of $\rho_0$ is obtained  by integrating the matter density distribution equivalent to the mass number of the daughter nucleus.\\

In equation (\ref{i1}), the density-dependence factor $F(\rho,E_\alpha)$ is given as \cite{deng2017,gontchar2010}:
\begin{equation}
F(\rho_1,\rho_2) = C \left[ 1 + \alpha e^{-\beta (\rho_1+\rho_2)}- \gamma (\rho_1+\rho_2) \right] .
\end{equation}

The parameters of the interaction viz. $C$, $\alpha$, $\beta$, $\gamma$ were determined through reproducing the saturation properties of normal nuclear matter within Hartree-Fock calculations \cite{khoa1997}. The  density-dependent $\mathrm{NN}$ interactions used in this paper is the DDM3Y1 parametrizations. The parameters $C$, $\alpha$, $\beta$, $\gamma$ corresponding to the DDM3Y1 parametrizations are given in Table \ref{tabDD}.
\begin{table}[!ht]
	\centering
	\caption{The parameters of the various density-dependent $\mathrm{N N}$ interactions used in this work \cite{khoa1997,zhang2009,gontchar2010}.}
	\begin{tabular}{lccccc}
		\hline \hline \\
		Interaction & Label & $C$ & $\alpha$ & $\beta$ & $\gamma$ \\ 
		\hline \hline \\
		D-independent & DD0 & 1 & 0 & 0 & 0 \\
		\\
		DDM3Y1 (Reid) & DD1& 0.2843 & 3.6391 & 2.9605 & 0.0000 \\ 
		\\
		DDM3Y1 (Paris) & DD1 & 0.2963 & 3.7231 & 3.7384 & 0.0000 \\ 
		\\
		\hline \hline \\
	\end{tabular} 
	\label{tabDD}
\end{table} 

The popular choices for nucleon-nucleon interactions in the double folding model have often been the M3Y interactions. The M3Y interactions were constructed to reproduce the G-matrix elements of both the Paris (M3Y-Paris) and Reid (M3Y-Reid) $\mathrm{NN}$ interactions in an oscillator basis \cite{zhang2009}. They are given by:
\begin{equation}
v^{M3Y-Paris}(s,E_{\alpha}) = 11062 \frac{e^{-4 s}}{4 s} - 2537.5 \frac{e^{-2.5 s}}{2.5 s} + J^{P}_{00} (E_{\alpha}) \delta(\textbf{s}) 
\end{equation}
and
\begin{equation}
v^{M3Y-Reid}(s,E_{\alpha}) = 7999 \frac{e^{-4 s}}{4 s} - 2134 \frac{e^{-2.5 s}}{2.5 s} + J^{R}_{00} (E_{\alpha}) \delta(\textbf{s})
\end{equation}
respectively. In this study, the effective nucleon-nucleon interactions derived from relativistic mean field (RMF) theory Lagrangian, with different parametrizations are also employed. Following Ref. \cite{singh2010}, the effective nucleon-nucleon interaction, derived from relativistic mean field Lagrangian is given by the sum of the scalar ($\sigma$) and vector ($\omega, \rho$) parts of the meson fields. That is,
\begin{align}
v_{eff} (s) &= V_{\omega} + V_{\sigma} + V_{\rho} \nonumber \\
& = \frac{g_{\omega}^2}{4 \pi} \frac{e^{-m_{\omega}s}}{s} - \frac{g_{\sigma}^2}{4 \pi} \frac{e^{-m_{\sigma}s}}{s} + \frac{g_{\rho}^2}{4 \pi} \frac{e^{-m_{\rho}s}}{s}   + J_{00} (E) \delta(s),
\end{align}
where $g_i$ and $m_i$ $(i=\omega, \sigma, \rho )$ are the coupling constants and meson masses, respectively, and the last term is the exchange contribution. Different parameters of the RMF effective $\mathrm{NN}$ interaction have been employed in this work viz. R3Y-L1, R3Y-W, R3Y-Z, and R3Y-HS parametrizations. They are given, respectively, as \cite{singh2010}: 
\begin{align}
v^{R3Y-L1}(s,E_{\alpha}) &= 9967.88 \frac{e^{-3.968 s}}{4 s} - 6660.95 \frac{e^{-2.787 s}}{4 s} + J^{R}_{00} (E_{\alpha}) \delta(\textbf{s}), \\
v^{R3Y-W}(s,E_{\alpha}) &= 8550.74 \frac{e^{-3.968 s}}{4 s} - 5750.24 \frac{e^{-2.787 s}}{4 s} + J^{R}_{00} (E_{\alpha}) \delta(\textbf{s}), \\
v^{R3Y-Z}(s,E_{\alpha}) &= 12008.98 \frac{e^{-3.9528 s}}{4 s} - 7861.80 \frac{e^{-2.7939 s}}{4 s} + J^{R}_{00} (E_{\alpha}) \delta(\textbf{s}),\\
v^{R3Y-HS}(s,E_{\alpha}) &= 11956.94 \frac{e^{-3.968 s}}{4 s} - 6882.64 \frac{e^{-2.6352 s}}{4 s} \nonumber \\
&+ 4099.06 \frac{e^{-3.902 s}}{4 s} + J^{P}_{00} (E_{\alpha}) \delta(\textbf{s}) .
\end{align}
A complete description of the R3Y interactions is provided in Ref. \cite{singh2010}. The zero-range exchange terms are given by
\begin{equation}
J^{R}_{00} (E_{\alpha}) = -276 (1 - 0.005 E_{\alpha}/A_{\alpha}) \; \mathrm{MeV} \: \mathrm{fm}^3 
\end{equation}
and
\begin{equation}
J^{P}_{00} (E_{\alpha}) = -590 (1 - 0.002 E_{\alpha}/A_{\alpha}) \; \mathrm{MeV} \: \mathrm{fm}^3 .
\end{equation}
Here $E_{\alpha} = Q_{\alpha} A_1/A$, $Q_{\alpha}$ denotes the energy released in the alpha decay process, and $A$ is the mass number of the parent nucleus. The quantization factor $\lambda$ in equation (\ref{veff}) is determined though the Bohr-Sommerfeld quantization and Wildermuth rule \cite{ghorbani2020,maroufi2019,christel2018}
\begin{equation}
\int_{r_1}^{r_2} \sqrt{\frac{2 \mu}{\hbar} \left[ Q_{\alpha} - V_{eff}(R) \right]} \: d R = \left( G - \ell + 1 \right) \frac{\pi}{2} ,
\end{equation}
where the global quantum number, $G$, is given for $\alpha-$decay process as
\begin{equation}
G_{\alpha} = \left\lbrace \begin{array}{ll}
18 & N \leq 82 \\ 
20 & 82< N \leq 126 \\ 
22 & N>126
\end{array}    \right. .
\end{equation}

The following formula is then used to calculate the $\alpha-$decay half-life \cite{ghorbani2020}:
\begin{equation}
T_{1/2} = \frac{\ln 2}{\nu P_\alpha P} ,
\end{equation}
where the assault frequency $\nu$ is determined using the WKB approximation \cite{maroufi2019polonica}
\begin{equation}
\nu = \frac{\hbar}{2 \mu} \left[ \int_{r_1}^{r_2} \frac{d R}{\sqrt{\frac{2 \mu}{\hbar^2} \left| Q - V_{eff}(R) \right| }} \right]^{-1} 
\end{equation}
and the tunneling probability $P$ is calculated via
\begin{equation}
P = \left( 1 + e^q \right)^{-1} ,
\end{equation}
and
\begin{equation}
q = \frac{\sqrt{8 \mu}}{\hbar} \int_{r_2}^{r_3} \sqrt{V_{eff}(R)-Q} \: d R ,
\end{equation}
$r_i (i=1,2,3)$ are the three turning points, and the pre-formation probability $P_\alpha$ is computed here using the empirical formula \cite{maroufi2019polonica}:
\begin{equation}
\log P_{\alpha} = s \sqrt{\mu Z_1 Z_2} + b ,
\label{eqpreform}
\end{equation}
where $a=-0.052$ and $b = 0.69$ for even-even nuclei. For odd-A nuclei, $b = 0.6$. 

\section{Results and Discussions}
\label{results}
\begin{figure}[!h]
	\centering
	\begin{subfigure}[b]{0.45\textwidth}
		\centering
		\includegraphics[width=\textwidth]{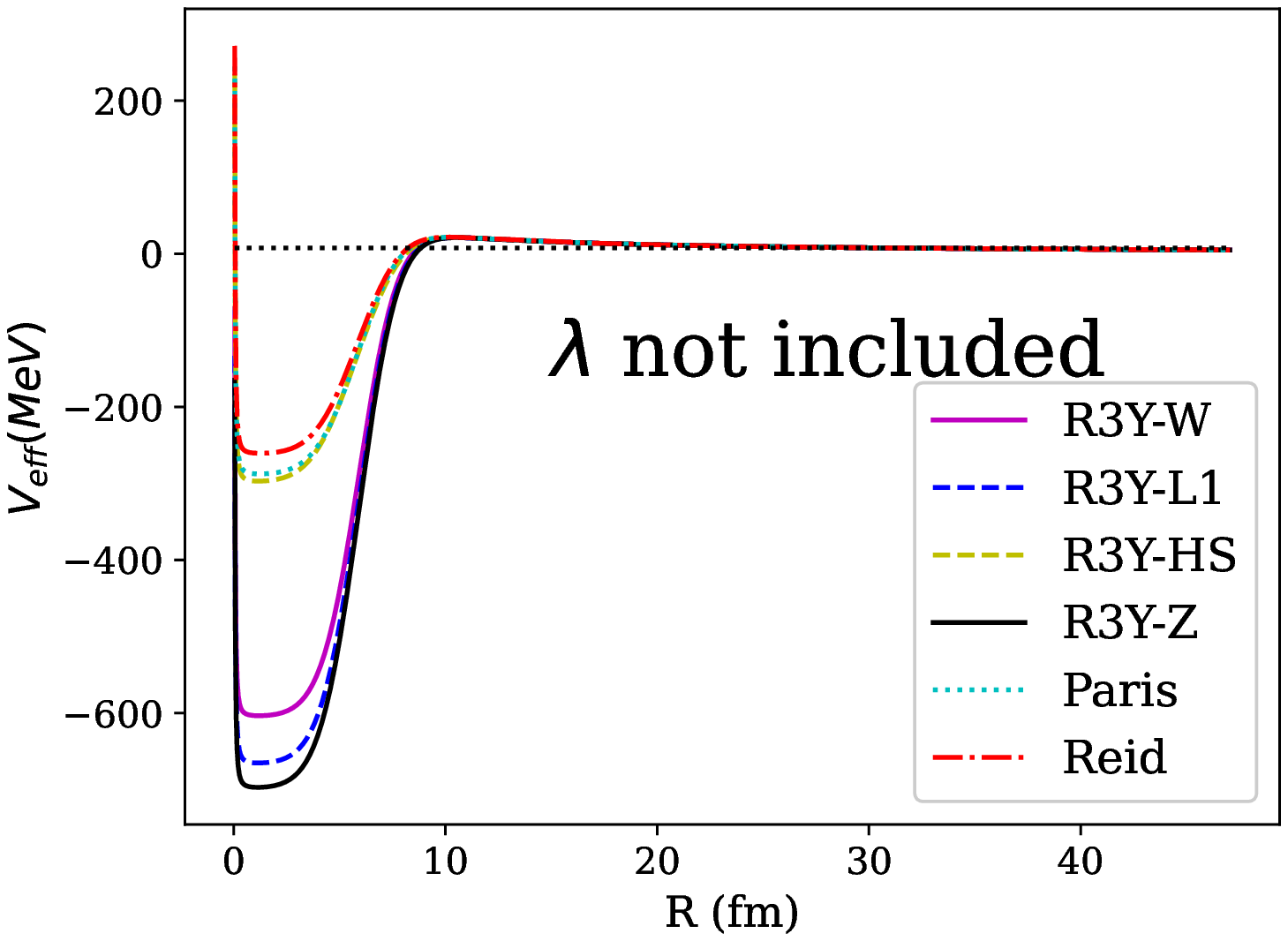}
		\caption{}
		\label{DD0no}
	\end{subfigure}
	\begin{subfigure}[b]{0.45\textwidth}
		\centering
		\includegraphics[width=\textwidth]{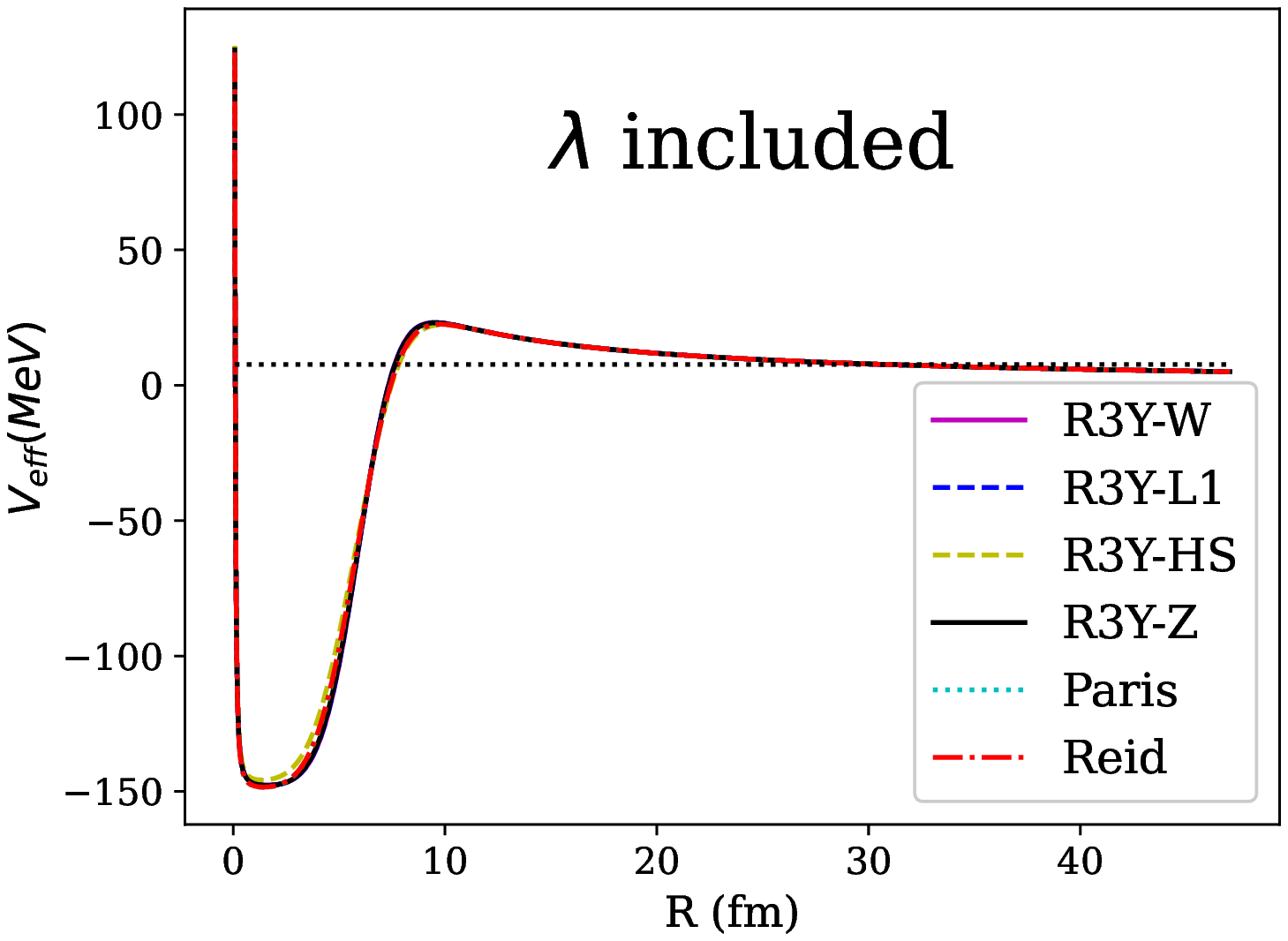}	
		\caption{}
		\label{DD0yes}
	\end{subfigure}
	\caption{Plot of the effective alpha-nucleus potential $V_{eff}$ for $^{190}\mathrm{Po}$ using density-independent (DD0) R3Y-W, R3Y-L1, R3Y-HS, R3Y-Z, M3Y-Paris, and M3Y-Reid  interactions. (a) quantization factor not applied and (b) quantization factor included.}
	\label{potentialDD0}
\end{figure}

Here the results of the calculations using the theory described above are presented and discussed. In the calculations, both density-independent (DD0) and density-dependent interactions (DDM3Y) were used. The experimental input data have been extracted from the NUBASE2020 database \cite{wang2021,huang2021,kondev2021}. In the calculations of the double folding potentials, the R3Y interactions with the different parametrizations (R3Y-HS, R3Y-L1, R3Y-W, and R3Y-Z) have been used. The calculations using the M3Y interactions are included for comparison with the R3Y interactions. In Figure \ref{potentialDD0}, the plots of the effective alpha-nucleus interactions (equation (\ref{veff})) using density-independent (DD0) R3Y-W, R3Y-L1, R3Y-HS, R3Y-Z, M3Y-Paris, and M3Y-Reid  interactions are shown. The quantization factor ($\lambda$) is not included in Figure \ref{DD0no}, whereas it is used in Figure \ref{DD0yes}. When the quantization factor is not included, the R3Y-Z can be seen to give the strongest potential while the M3Y-Reid gives the weakest potential. However when the quantization factor is used, only a slight difference is observed in the strengths of the potentials for the different models. The quantization factor has the most effect on the R3Y-Z potential, by drastically reducing the strength of the potential. The black dots in Figures \ref{DD0no} and \ref{DD0yes} indicate the $Q_{\alpha}$ values.  \\

In order to give a quantitative comparison between the theoretically calculated results and the experimental data, the root mean square standard deviation $(\sigma)$ has been computed for the different models. The following formula was used to compute the standard deviation: \cite{yahya2020}:
\begin{equation}
\sigma = \sqrt{\frac{1}{N} \sum_{i=1}^{N} \left[ \left(  \log_{10} T_{1/2,i}^{ \mathrm{Theory}}   - \log_{10} T_{1/2,i}^{ \mathrm{Expt}} \right)^2  \right] } .
\label{sigma}
\end{equation}
Here $T_{1/2,i}^{\mathrm{Expt}}$ are the experimental half-lives while $T_{1/2,i}^{\mathrm{Theory}}$ are the theoretical half-lives. \\

The calculated $\alpha-$decay half-lives for the $33$ Polonium  $(\mathrm{Po}) $ isotopes using the double folding model with density-independent interactions (i.e. DD0) are shown in Table \ref{tab:halflifeDD0_no_preform}. Here the preformation factor $P_{\alpha}$ is taken to be one. The first three columns show, respectively, the mass number ($A$), experimental 
$Q_{\alpha}$ values, and the logarithm of the experimental $\alpha$-decay half-lives. The fourth to ninth columns show the results using the M3Y-Paris, M3Y-Reid, R3Y-HS, R3Y-L1, R3Y-W, and R3Y-Z parameters, respectively. The last row of the Table shows the calculated standard deviation values $(\sigma)$ for the various models. The $\sigma$ for the M3Y-Paris, M3Y-Reid, R3Y-HS, R3Y-L1, R3Y-W, and R3Y-Z models are $0.8044, \, 0.8099,\, 0.7807, \, 0.5729$, and $0.5595$, respectively. The R3Y models have lower $\sigma$ than the M3Y models, which suggests that the R3Y models give better descriptions of the $\alpha-$decay half-lives of the Polonium isotopes than the M3Y models. \\
\begin{table}[!h]
	\centering
	\caption{Calculated $\alpha$-decay half-lives, $\log \left[ T_{1/2} (s) \right]$, of $\mathrm{Po}$ isotopes $(Z=84)$ using density-independent (DD0) M3Y and R3Y interactions and setting $P_{\alpha}=1$.}
	\begin{tabular}{ccccccccc}
		\hline \hline \\
		& & \multicolumn{7}{c}{$\log \left[ T_{1/2} (s) \right]$} \\
		\cline{3-9} 
		A & $Q_{\alpha}$ & Expt.  & M3Y-Paris & M3Y-Reid & R3Y-HS & R3Y-L1  & R3Y-W & R3Y-Z   \\
		\hline \hline \\
		186  &    8.5012  &   -4.3980  &   -5.4067  &   -5.3986  &   -5.5411  &   -5.1470  &   -5.1298  &   -5.1767   \\ 
		187  &    7.9789  &   -2.8540  &   -3.6645  &   -3.6744  &   -3.8078  &   -3.4051  &   -3.3669  &   -3.4357   \\ 
		188  &    8.0823  &   -3.5610  &   -4.2642  &   -4.2736  &   -4.3805  &   -3.9779  &   -3.9813  &   -4.0288   \\ 
		189  &    7.6943  &   -2.4200  &   -2.8340  &   -2.8430  &   -2.9770  &   -2.5668  &   -2.5340  &   -2.5976   \\ 
		190  &    7.6933  &   -2.6090  &   -3.1242  &   -3.1327  &   -3.1493  &   -2.8523  &   -2.8201  &   -2.8689   \\ 
		191  &    7.8223  &   -1.6580  &   -3.5370  &   -3.5456  &   -3.5871  &   -3.2693  &   -3.2521  &   -3.2844   \\ 
		192  &    7.3196  &   -1.4920  &   -1.9469  &   -1.9498  &   -1.8905  &   -1.6678  &   -1.6500  &   -1.6987   \\ 
		193  &    7.0938  &   -0.4320  &   -1.1873  &   -1.1942  &   -1.0868  &   -0.9034  &   -0.8855  &   -0.9345   \\ 
		194  &    6.9871  &   -0.4070  &   -0.8235  &   -0.8301  &   -0.7033  &   -0.5378  &   -0.5157  &   -0.5689   \\ 
		195  &    6.7497  &    0.6670  &    0.0367  &    0.0306  &    0.1934  &    0.3272  &    0.3455  &    0.2959   \\ 
		196  &    6.6582  &    0.7450  &    0.3714  &    0.3654  &    0.5402  &    0.6635  &    0.6817  &    0.6319   \\ 
		197  &    6.4113  &    2.0800  &    1.3394  &    1.3337  &    1.5323  &    1.6357  &    1.6541  &    1.6038   \\ 
		198  &    6.3097  &    2.0260  &    1.7461  &    1.7403  &    1.9388  &    2.0435  &    2.0620  &    2.0113   \\ 
		199  &    6.0743  &    3.6400  &    2.7518  &    2.7463  &    2.9546  &    3.0513  &    3.0698  &    3.0192   \\ 
		200  &    5.9816  &    3.7900  &    3.1562  &    3.1504  &    3.3567  &    3.4558  &    3.4743  &    3.4237   \\ 
		201  &    5.7993  &    4.7600  &    3.9990  &    3.9934  &    4.1857  &    4.2972  &    4.3160  &    4.2657   \\ 
		202  &    5.7010  &    5.1500  &    4.4653  &    4.4593  &    4.6398  &    4.7626  &    4.7808  &    4.7307   \\ 
		203  &    5.4960  &    6.3000  &    5.7751  &    5.7683  &    5.9027  &    6.0713  &    6.0908  &    6.0396   \\ 
		204  &    5.4849  &    6.2800  &    5.5427  &    5.5365  &    5.6790  &    5.8348  &    5.8530  &    5.8032   \\ 
		205  &    5.3247  &    7.1800  &    6.3955  &    6.3892  &    6.4928  &    6.6820  &    6.7000  &    6.6509   \\ 
		206  &    5.3270  &    7.1500  &    6.3681  &    6.3615  &    6.4666  &    6.6538  &    6.6718  &    6.6228   \\ 
		207  &    5.2159  &    8.0000  &    6.9800  &    6.9725  &    7.0472  &    7.2603  &    7.2781  &    7.2297   \\ 
		208  &    5.2157  &    7.9610  &    6.9662  &    6.9592  &    7.0341  &    7.2459  &    7.2632  &    7.2174   \\ 
		209  &    4.9792  &    9.5070  &    8.6379  &    8.6304  &    8.6136  &    8.9069  &    8.9190  &    8.8773   \\ 
		210  &    5.4075  &    7.0780  &    5.8673  &    5.8609  &    5.9888  &    6.1545  &    6.1724  &    6.1233   \\ 
		211  &    7.5946  &   -0.2870  &   -2.0614  &   -2.0684  &   -2.2014  &   -1.7798  &   -1.7619  &   -1.8096   \\ 
		212  &    8.9542  &   -6.5240  &   -7.1707  &   -7.1783  &   -7.3548  &   -6.9657  &   -6.9520  &   -6.9799   \\ 
		213  &    8.5361  &   -5.4290  &   -6.1067  &   -6.1075  &   -6.2109  &   -5.8874  &   -5.8731  &   -5.9122   \\ 
		214  &    7.8335  &   -3.7840  &   -4.1262  &   -4.1323  &   -4.0893  &   -3.8779  &   -3.8620  &   -3.9052   \\ 
		215  &    7.5263  &   -2.7490  &   -3.1842  &   -3.1898  &   -3.0869  &   -2.9226  &   -2.9061  &   -2.9511   \\ 
		216  &    6.9063  &   -0.8390  &   -1.0826  &   -1.0937  &   -0.8790  &   -0.7942  &   -0.7763  &   -0.8251   \\ 
		217  &    6.6621  &    0.1800  &   -0.1825  &   -0.1871  &    0.0510  &    0.1151  &    0.1334  &    0.0834   \\ 
		218  &    6.1148  &    2.2690  &    2.0579  &    2.0532  &    2.3217  &    2.3686  &    2.3875  &    2.3357   \\ 
		
		\hline \hline	\\
		$\sigma$  & & & 0.8044 & 0.8099 & 0.7807 & 0.5729 & 0.5595 & 0.5950	\\
		\hline \hline	
	\end{tabular}
	\label{tab:halflifeDD0_no_preform}
\end{table}
\begin{table}[!h]
	\centering
	\caption{Calculated $\alpha$-decay half-lives, $\log \left[ T_{1/2} (s) \right]$, of $\mathrm{Po}$ isotopes $(Z=84)$ using density-independent (DD0) interactions and including the pre-formation factor $P_{\alpha}$}
	\begin{tabular}{ccccccccc}
		\hline \hline \\
		& & \multicolumn{5}{c}{$\log \left[ T_{1/2} (s) \right]$} & \\
		\cline{3-7} 
		A & $Q_{\alpha}$ & Expt. & R3Y-HS & R3Y-L1  & R3Y-W & R3Y-Z & $\log P_{\alpha}$   \\
		\hline \hline \\		
		
		186  &    8.5012  &   -4.3980  &   -4.9137  &   -4.5195  &   -4.5024  &   -4.5493  &   -0.6274  \\ 
		187  &    7.9789  &   -2.8540  &   -3.0902  &   -2.6876  &   -2.6494  &   -2.7182  &   -0.7175  \\ 
		188  &    8.0823  &   -3.5610  &   -3.7529  &   -3.3503  &   -3.3537  &   -3.4012  &   -0.6276  \\ 
		189  &    7.6943  &   -2.4200  &   -2.2593  &   -1.8492  &   -1.8163  &   -1.8799  &   -0.7177  \\ 
		190  &    7.6933  &   -2.6090  &   -2.5216  &   -2.2246  &   -2.1923  &   -2.2411  &   -0.6278  \\ 
		191  &    7.8223  &   -1.6580  &   -2.8693  &   -2.5515  &   -2.5343  &   -2.5666  &   -0.7178  \\ 
		192  &    7.3196  &   -1.4920  &   -1.2626  &   -1.0399  &   -1.0221  &   -1.0708  &   -0.6279  \\ 
		193  &    7.0938  &   -0.4320  &   -0.3689  &   -0.1855  &   -0.1675  &   -0.2165  &   -0.7180  \\ 
		194  &    6.9871  &   -0.4070  &   -0.0753  &    0.0903  &    0.1123  &    0.0592  &   -0.6280  \\ 
		195  &    6.7497  &    0.6670  &    0.9115  &    1.0453  &    1.0636  &    1.0140  &   -0.7181  \\ 
		196  &    6.6582  &    0.7450  &    1.1684  &    1.2916  &    1.3099  &    1.2601  &   -0.6282  \\ 
		197  &    6.4113  &    2.0800  &    2.2506  &    2.3539  &    2.3723  &    2.3220  &   -0.7183  \\ 
		198  &    6.3097  &    2.0260  &    2.5671  &    2.6719  &    2.6903  &    2.6396  &   -0.6283  \\ 
		199  &    6.0743  &    3.6400  &    3.6730  &    3.7697  &    3.7882  &    3.7376  &   -0.7184  \\ 
		200  &    5.9816  &    3.7900  &    3.9852  &    4.0843  &    4.1028  &    4.0522  &   -0.6285  \\ 
		201  &    5.7993  &    4.7600  &    4.9042  &    5.0157  &    5.0345  &    4.9842  &   -0.7185  \\ 
		202  &    5.7010  &    5.1500  &    5.2684  &    5.3913  &    5.4094  &    5.3593  &   -0.6286  \\ 
		203  &    5.4960  &    6.3000  &    6.6214  &    6.7900  &    6.8095  &    6.7583  &   -0.7187  \\ 
		204  &    5.4849  &    6.2800  &    6.3078  &    6.4635  &    6.4818  &    6.4320  &   -0.6287  \\ 
		205  &    5.3247  &    7.1800  &    7.2116  &    7.4008  &    7.4188  &    7.3697  &   -0.7188  \\ 
		206  &    5.3270  &    7.1500  &    7.0955  &    7.2827  &    7.3007  &    7.2517  &   -0.6289  \\ 
		207  &    5.2159  &    8.0000  &    7.7661  &    7.9793  &    7.9970  &    7.9486  &   -0.7189  \\ 
		208  &    5.2157  &    7.9610  &    7.6631  &    7.8749  &    7.8922  &    7.8464  &   -0.6290  \\ 
		209  &    4.9792  &    9.5070  &    9.3326  &    9.6260  &    9.6380  &    9.5963  &   -0.7190  \\ 
		210  &    5.4075  &    7.0780  &    6.6179  &    6.7836  &    6.8015  &    6.7524  &   -0.6291  \\ 
		211  &    7.5946  &   -0.2870  &   -1.4823  &   -1.0606  &   -1.0427  &   -1.0905  &   -0.7192  \\ 
		212  &    8.9542  &   -6.5240  &   -6.7256  &   -6.3365  &   -6.3228  &   -6.3506  &   -0.6292  \\ 
		213  &    8.5361  &   -5.4290  &   -5.4916  &   -5.1681  &   -5.1538  &   -5.1929  &   -0.7193  \\ 
		214  &    7.8335  &   -3.7840  &   -3.4600  &   -3.2485  &   -3.2327  &   -3.2758  &   -0.6293  \\ 
		215  &    7.5263  &   -2.7490  &   -2.3675  &   -2.2032  &   -2.1867  &   -2.2317  &   -0.7194  \\ 
		216  &    6.9063  &   -0.8390  &   -0.2495  &   -0.1647  &   -0.1468  &   -0.1957  &   -0.6295  \\ 
		217  &    6.6621  &    0.1800  &    0.7705  &    0.8346  &    0.8529  &    0.8029  &   -0.7195  \\ 
		218  &    6.1148  &    2.2690  &    2.9513  &    2.9981  &    3.0171  &    2.9652  &   -0.6296  \\  	
		\hline \hline		
	\end{tabular}
	\label{tab:halflifeDD0_yes_preform}
\end{table}
\begin{table}[!h]
	\centering
	\caption{Calculated $\alpha$-decay half-lives, $\log \left[ T_{1/2} (s) \right]$, of $\mathrm{Po}$ isotopes $(Z=84)$ using density-dependent (DDM3Y) interactions and including the pre-formation factor $P_{\alpha}$}
	\begin{tabular}{ccccccccc}
		\hline \hline \\
		& & \multicolumn{5}{c}{$\log \left[ T_{1/2} (s) \right]$} & \\
		\cline{3-7} 
		A & $Q_{\alpha}$ & Expt. & R3Y-HS & R3Y-L1  & R3Y-W & R3Y-Z & $\log P_{\alpha}$   \\
		\hline \hline \\			
		186  &    8.5012  &   -4.3980  &   -4.5453  &   -4.7343  &   -4.7061  &   -4.7643  &   -0.6274  \\  
		187  &    7.9789  &   -2.8540  &   -2.7375  &   -2.9087  &   -2.8805  &   -2.9267  &   -0.7175  \\  
		188  &    8.0823  &   -3.5610  &   -3.4126  &   -3.5880  &   -3.5600  &   -3.6193  &   -0.6276  \\  
		189  &    7.6943  &   -2.4200  &   -1.9107  &   -2.0720  &   -2.0431  &   -2.0955  &   -0.7177  \\  
		190  &    7.6933  &   -2.6090  &   -2.3080  &   -2.4443  &   -2.4158  &   -2.4764  &   -0.6278  \\  
		191  &    7.8223  &   -1.6580  &   -2.6032  &   -2.7693  &   -2.7412  &   -2.8009  &   -0.7178  \\  
		192  &    7.3196  &   -1.4920  &   -1.1169  &   -1.2627  &   -1.2339  &   -1.2957  &   -0.6279  \\  
		193  &    7.0938  &   -0.4320  &   -0.1526  &   -0.4102  &   -0.3812  &   -0.4440  &   -0.7180  \\  
		194  &    6.9871  &   -0.4070  &    0.1346  &   -0.1342  &   -0.1058  &   -0.1684  &   -0.6280  \\  
		195  &    6.7497  &    0.6670  &    0.9515  &    0.8178  &    0.8476  &    0.7814  &   -0.7181  \\  
		196  &    6.6582  &    0.7450  &    1.3903  &    1.0634  &    1.0933  &    1.0261  &   -0.6282  \\  
		197  &    6.4113  &    2.0800  &    2.2417  &    2.1237  &    2.1540  &    2.0844  &   -0.7183  \\  
		198  &    6.3097  &    2.0260  &    2.5576  &    2.4411  &    2.4714  &    2.4055  &   -0.6283  \\  
		199  &    6.0743  &    3.6400  &    3.6532  &    3.5376  &    3.5676  &    3.5017  &   -0.7184  \\  
		200  &    5.9816  &    3.7900  &    3.9680  &    3.8520  &    3.8825  &    3.8161  &   -0.6285  \\  
		201  &    5.7993  &    4.7600  &    4.9028  &    4.7805  &    4.8143  &    4.7439  &   -0.7185  \\  
		202  &    5.7010  &    5.1500  &    5.2811  &    5.1589  &    5.1894  &    5.1233  &   -0.6286  \\  
		203  &    5.4960  &    6.3000  &    6.6933  &    6.5548  &    6.5854  &    6.5192  &   -0.7187  \\  
		204  &    5.4849  &    6.2800  &    6.3641  &    6.2324  &    6.2625  &    6.1974  &   -0.6287  \\  
		205  &    5.3247  &    7.1800  &    7.3128  &    7.1707  &    7.2009  &    7.1369  &   -0.7188  \\  
		206  &    5.3270  &    7.1500  &    7.1946  &    7.0534  &    7.0829  &    7.0192  &   -0.6289  \\  
		207  &    5.2159  &    8.0000  &    7.9006  &    7.7514  &    7.7806  &    7.7179  &   -0.7189  \\  
		208  &    5.2157  &    7.9610  &    7.7963  &    7.6475  &    7.6766  &    7.6140  &   -0.6290  \\  
		209  &    4.9792  &    9.5070  &    9.5689  &    9.3973  &    9.4263  &    9.3653  &   -0.7190  \\  
		210  &    5.4075  &    7.0780  &    6.6895  &    6.5548  &    6.5843  &    6.5203  &   -0.6291  \\  
		211  &    7.5946  &   -0.2870  &   -1.1383  &   -1.2869  &   -1.2578  &   -1.3202  &   -0.7192  \\  
		212  &    8.9542  &   -6.5240  &   -6.3515  &   -6.5192  &   -6.4972  &   -6.5430  &   -0.6292  \\  
		213  &    8.5361  &   -5.4290  &   -5.2023  &   -5.3571  &   -5.3304  &   -5.3826  &   -0.7193  \\  
		214  &    7.8335  &   -3.7840  &   -3.3201  &   -3.4498  &   -3.4241  &   -3.4791  &   -0.6293  \\  
		215  &    7.5263  &   -2.7490  &   -2.2922  &   -2.4099  &   -2.3831  &   -2.4409  &   -0.7194  \\  
		216  &    6.9063  &   -0.8390  &   -0.2873  &   -0.3826  &   -0.3536  &   -0.4172  &   -0.6295  \\  
		217  &    6.6621  &    0.1800  &    0.6948  &    0.6129  &    0.6427  &    0.5772  &   -0.7195  \\  
		218  &    6.1148  &    2.2690  &    2.8513  &    2.7702  &    2.8012  &    2.7327  &   -0.6296  \\  		
		\hline \hline		
	\end{tabular}
	\label{tab:halflifeDD1_yes_preform}
\end{table}

In Tables \ref{tab:halflifeDD0_yes_preform} and \ref{tab:halflifeDD1_yes_preform}, the results of the calculated $\alpha-$decay half-lives for the Polonium isotopes are shown using density-independent and density-dependent interactions, respectively. In both Tables, the pre-formation factor using equation (\ref{eqpreform}) is included. The fourth to seventh columns show the results using the R3Y-HS, R3Y-L1, R3Y-W, and R3Y-Z models, respectively. The last column shows the calculated pre-formation factor ($\log P_{\alpha}$). A physical inspection of the Tables indicate that the R3Y-models give very good descriptions of the $\alpha-$decay half-lives of the Polonium isotopes. Moreover, Table \ref{tab:rms} shows the results of the standard deviation $(\sigma)$ calculations using the data in Tables \ref{tab:halflifeDD0_yes_preform} and \ref{tab:halflifeDD1_yes_preform}. When density-independent model is used, the four R3Y models viz. R3Y-HS, R3Y-L1, R3Y-W and R3Y-Z have the respective standard deviation values $0.4278, 0.4328, 0.4440$, and $0.4159$. This confirms that all the R3Y models give very good descriptions of the $\alpha-$decay half-lives of the $33$ polonium isotopes. The R3Y-Z gives the lowest value of $\sigma$ while the R3Y-W gives the highest value. Furthermore, when the density-dependent interaction (DDM3Y) is used, the standard deviation values decrease for the four R3Y models. This shows the importance of using density-dependent interactions in the R3Y models. \\
\begin{table}[!ht]
	\centering
	\caption{The calculated root mean square standard deviations }
	\begin{tabular}{ccccc}
		\hline \hline \\
		& R3Y-HS & R3Y-L1 & R3Y-W & R3Y-Z   \\ \\
		\hline \hline \\
		DD0	& 0.4278 & 0.4328 & 0.4440 & 0.4159 \\ \\
		
		DDM3Y & 0.3970 & 0.3627 & 0.3651 & 0.3626 \\ \\
		
		\hline \hline \\ 
	\end{tabular} 
	\label{tab:rms}
\end{table}

\begin{figure}[!h]
	\centering
	\begin{subfigure}[b]{0.49\textwidth}
		\centering
		\includegraphics[width=\textwidth]{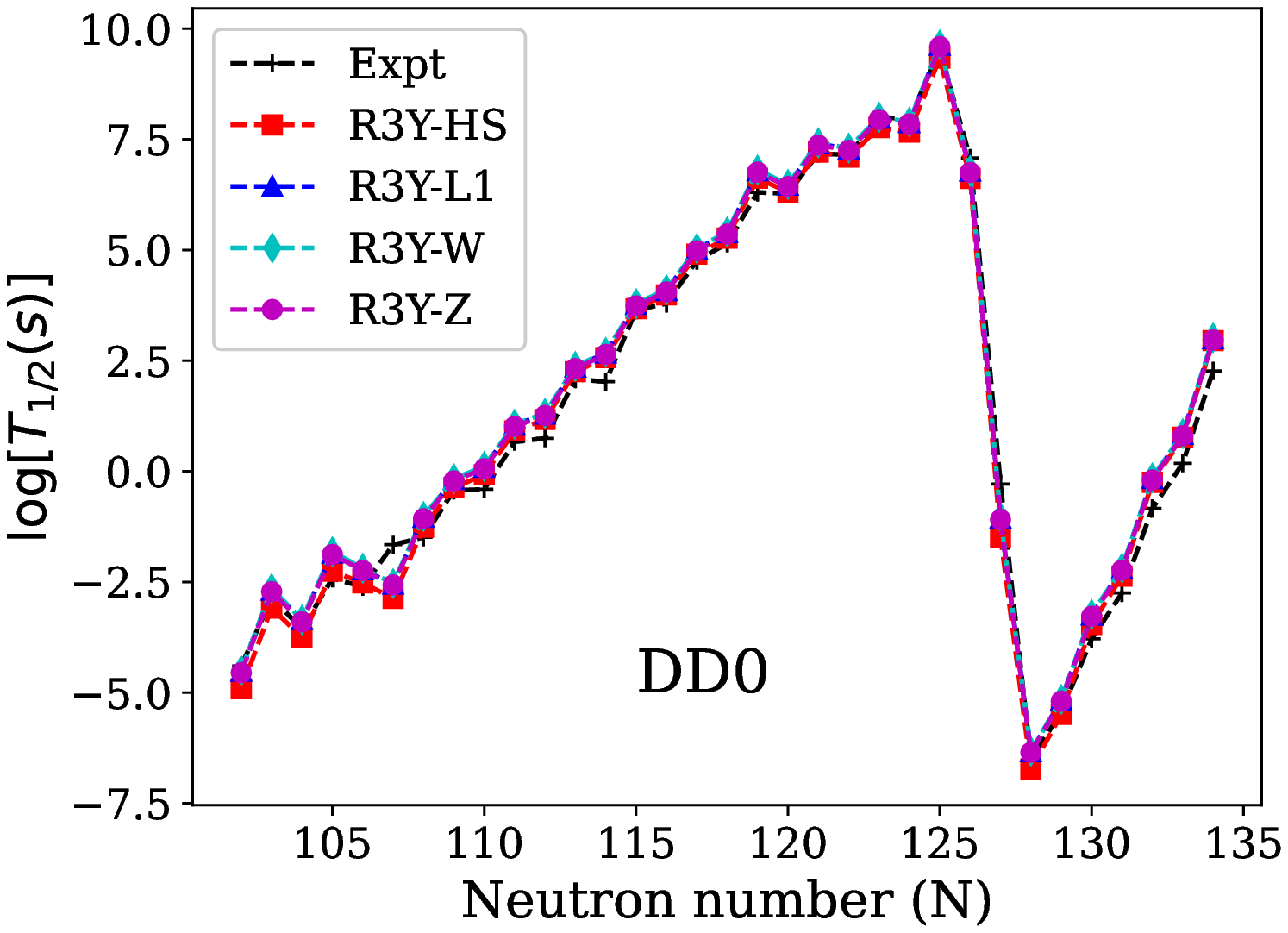}
		\caption{}
		\label{dd0}
	\end{subfigure}
	\begin{subfigure}[b]{0.49\textwidth}
		\centering
		\includegraphics[width=\textwidth]{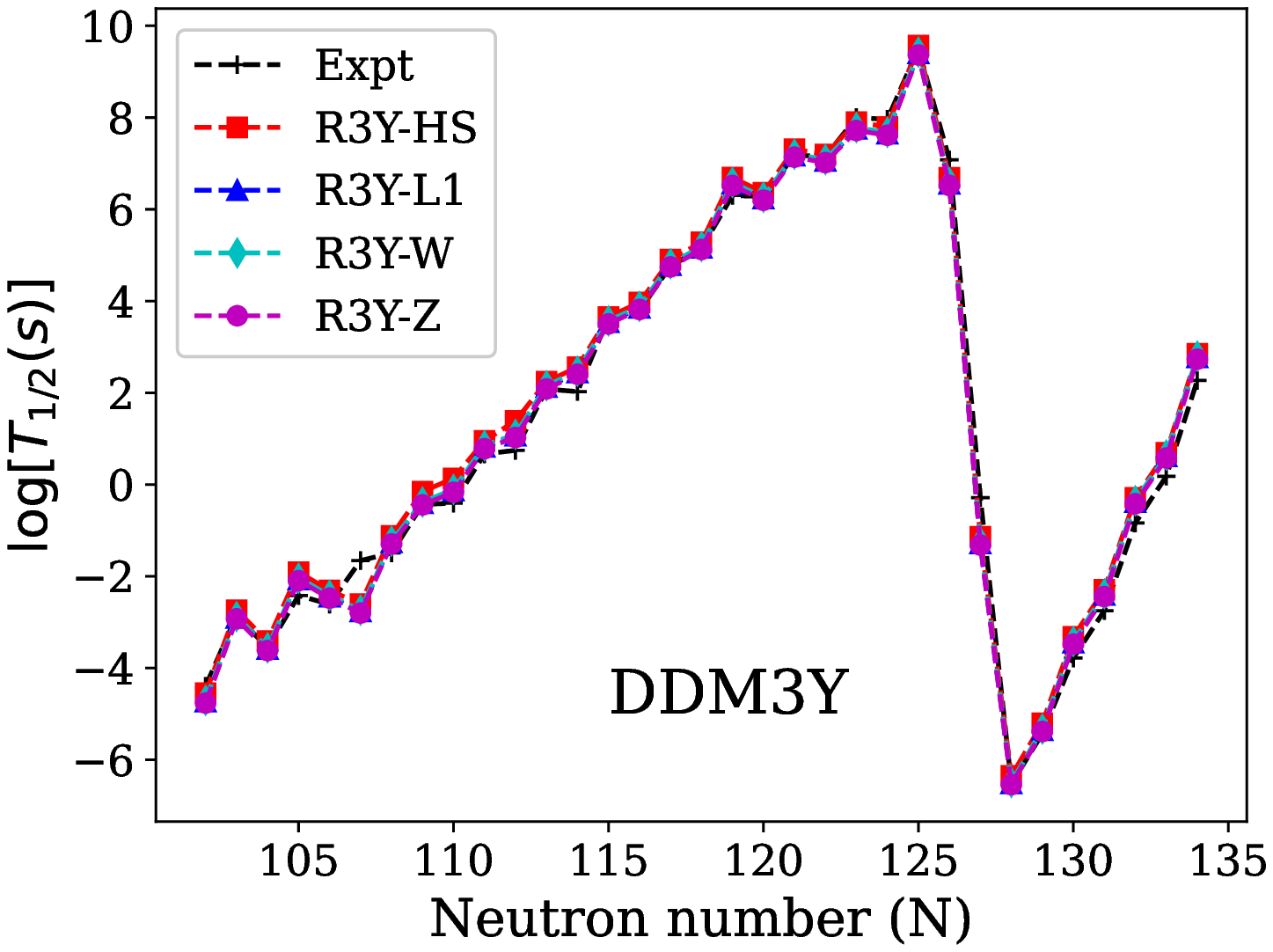}	
		\caption{}
		\label{ddm3y}
	\end{subfigure}
	\caption{Comparison of the calculated $\alpha-$decay half-lives of the $\mathrm{Po}$ isotopes between the theoretical	models and experiment. (a)  using density-independent (DD0) interactions (b) using density-dependent DDM3Y interactions.}
	\label{halflife}
\end{figure}

The plots of the calculated $\alpha-$decay half-lives $\log \left[ T_{1/2} (s) \right]$ against the neutron number using the four R3Y-models with the experimental half-lives are shown in Figure \ref{halflife}. The density-independent model (DD0) is shown in Figure \ref{dd0} while the density-dependent DDM3Y model is shown in Figure \ref{ddm3y}. The maximum value of the $\alpha-$decay half-lives is obtained at $N=125$ which corresponds to the parent nucleus $^{209} \mathrm{Po} $. The minimum value of the $\alpha-$decay half-lives is obtained at $N=128$ which corresponds to the daughter nucleus $^{208} \mathrm{Pb} $ with neutron number $N=126$. The maximum and minimum values are associated with the role of shell closure effects relative to the magicity (or near magicity) of the neutron number. A high half-life indicates the magicity of the parent nucleus, while a low half-life indicates the magicity of the daughter nucleus. Here the daughter nucleus that corresponds to the lowest half-life $\left( ^{208} \mathrm{Pb}  \right)$ has a neutron magic number $N=126$. \\

\begin{figure}[!h]
	\centering
	\begin{subfigure}[b]{0.49\textwidth}
		\centering
		\includegraphics[width=\textwidth]{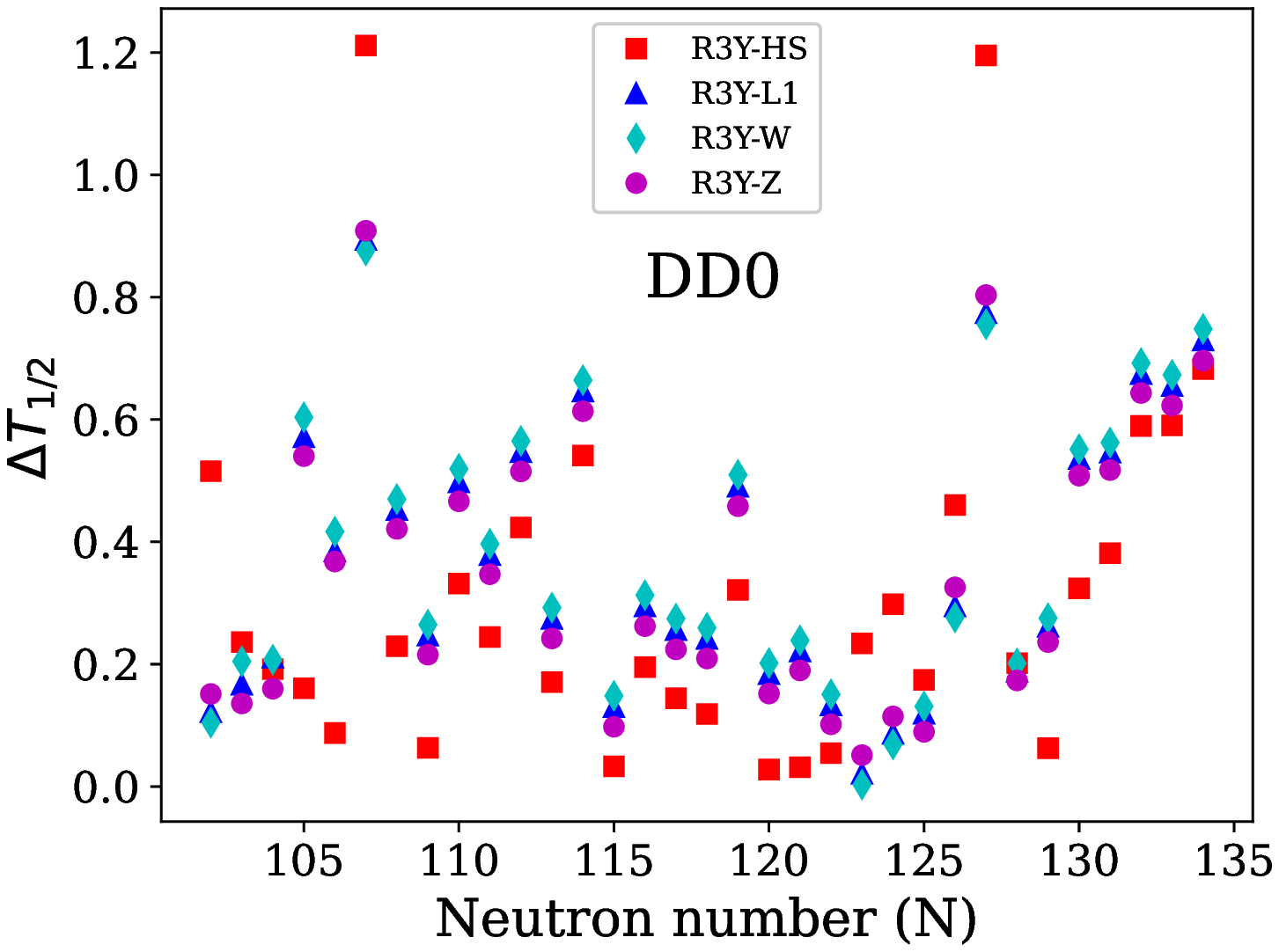}
		\caption{}
		\label{delta0}
	\end{subfigure}
	\begin{subfigure}[b]{0.49\textwidth}
		\centering
		\includegraphics[width=\textwidth]{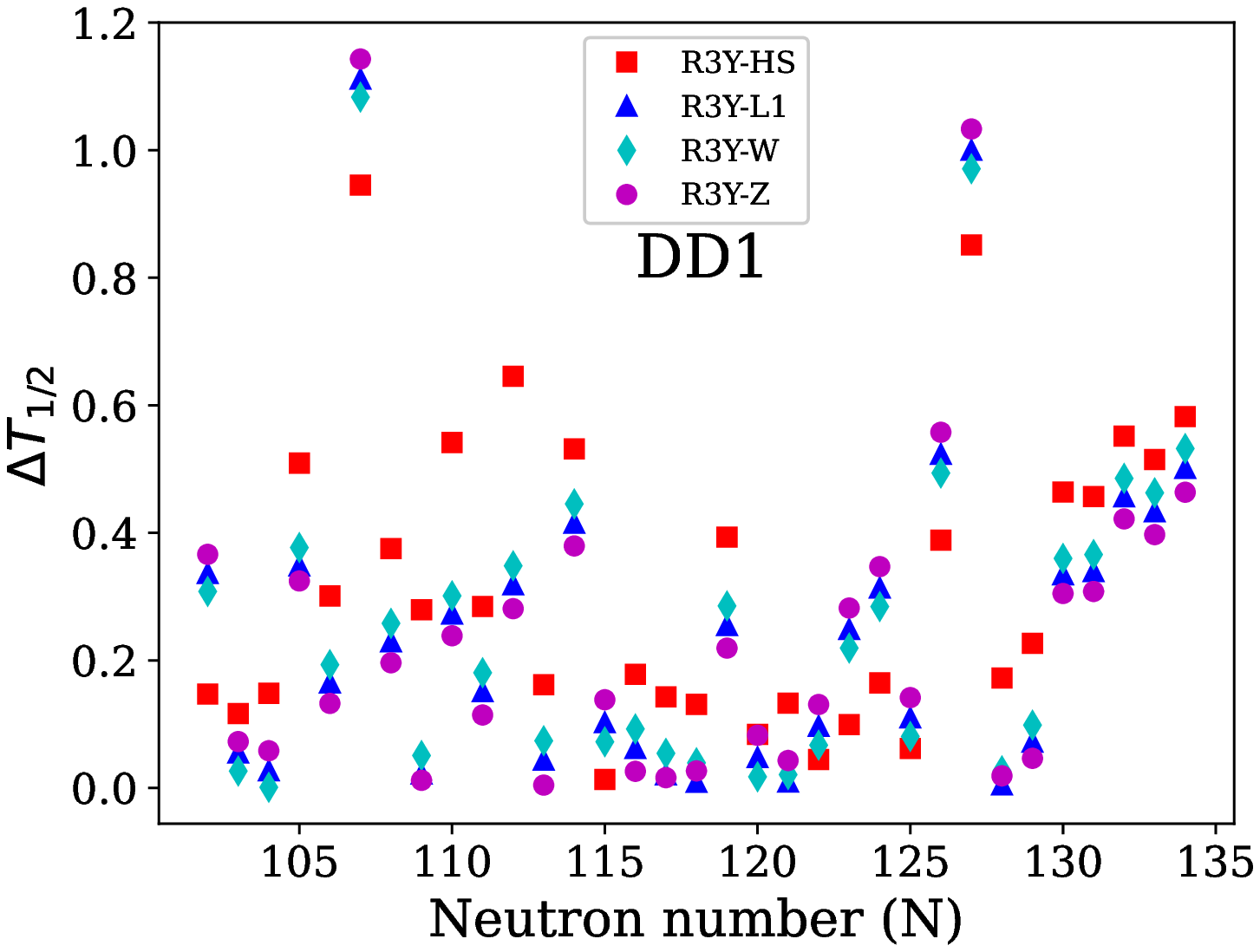}	
		\caption{}
		\label{delta1}
	\end{subfigure}
	\caption{Plot of the calculated $\Delta T_{1/2}$ against neutron number $(N)$ for the $\mathrm{Po}$ (a) using density-independent (DD0) interactions (b) using density-dependent DDM3Y interactions.}
	\label{DeltaT}
\end{figure}
\FloatBarrier

The difference between the experimental and theoretical $\alpha$-decay half-lives has also been calculated using the following formula \cite{akrawy2019,yahya2020}:
\begin{equation}
\Delta T_{1/2} =\left| \log_{10} \left[ T_{1/2}^{\mathrm{theor}} \right]  -  \log_{10} \left[ T_{1/2}^{\mathrm{expt}} \right] \right| .
\end{equation}
Figure \ref{DeltaT} shows the plots of $\Delta T_{1/2} $ against neutron number for the different models. In Figure \ref{delta0}, the computed $\Delta T_{1/2} $ using the density-independent models (DD0) are shown while Figure \ref{delta1} shows the results using the density-dependent DDM3Y models. In the two plots (Figure \ref{delta0} and Figure \ref{delta1}), most of the points are below $0.6$. This again confirms the accuracy of the use of the R3Y models to study the $\alpha-$decay half-lives of the Polonium isotopes. \\

\section{Conclusion}
\label{conclusion}	

The calculations of the $\alpha$-decay half-lives of some Polonium isotopes in the mass range $186-218$ have been carried out theoretically using the WKB semiclassical approximations and with the use of the Bohr-Sommerfeld quantization factor. The $\alpha-$nucleus potential is obtained using the double folding model, with the R3Y nucleon-nucleon effective interactions. The R3Y effective nucleon-nucleon interactions are derived from relativistic mean field theory Lagrangian. For comparison, the calcuations using the M3Y interactions were also included. When compared with experimental data, the results obtained using the R3Y models are found to be better than the results obtained using the M3Y-Reid and M3Y-Paris $\mathrm{NN}$ interactions. When density-dependent DDM3Y interactions are used in the R3Y models, the results are found to be better than using density-independent interactions, with the R3Y-Z giving the lowest deviation from experimental data. In general, when compared to experimental data, the R3Y models give maximum standard deviation value $\sigma = 0.4440$ when density-independent interaction is used and maximum $\sigma=0.3970$ when density-dependent interaction is employed. This shows the importance of using density-dependent interaction in the R3Y model. We conclude that the use of the R3Y effective $\mathrm{NN}$ interactions in the double folding model give very good descriptions of the alpha-decay half-lives of the Polonium isotopes.


\bibliographystyle{unsrt}
\bibliography{alphadecay}	

\end{document}